\pdfoutput=1
\documentclass[a4paper,11pt]{article}

\usepackage{jcappub}
\usepackage{lineno}
\usepackage{graphicx}
\usepackage{subfig}
\usepackage{booktabs}
\usepackage{amsmath}
\usepackage{mathtools}
\usepackage{xurl}
\usepackage{hyperref}
\usepackage{float}
\usepackage{placeins}
\usepackage{longtable}
\usepackage{multirow}

\arxivnumber{2609.00000}
\title{\boldmath Studies on the dark sector interaction from joint analysis of cosmological probes}

\author[1,2]{Jianfeng Meng}
\author[1]{Xiaofeng Yang$^*$}
\author[1,3]{Yunliang Ren}
\author[1,2]{Bohao Wang}
\author[1,2]{Jingze Li}
\author[1]{Kang Jiao}
\author[2]{Xiongwei Liu}
\affiliation[1]{\small School of Physics and Electronics, Henan University, Jinming Avenue, Kaifeng 450046, China}
\affiliation[2]{\small School of Physics and Astronomy, China West Normal University, No.1 Shida Road, Nanchong 637002, China}
\affiliation[3]{\small School of Physical Science and Technology, Xinjiang University, No.666 Shengli Road, Urumqi 830046, China}

\emailAdd{xfyang@henu.edu.cn}

\abstract{We test whether constraints on the nonlinear interaction $\xi$IDE are stable under different treatments of the Type Ia supernovae absolute calibration. \textit{Fermi} GRBs measurements and the Amati-relation parameters are fitted jointly with PantheonPlus SNe Ia, DESI DR2 BAO, and an updated cosmic-chronometer compilation. We compare the PantheonPlus-SH0ES route, which retains the SN absolute calibration, with the PantheonPlus-only route, in which the SN absolute magnitude is analytically marginalized. The GOLD GRB sample is adopted for the main analysis, while the FULL GRB sample is used to assess sample dependence. For the interaction parameter $\gamma\equiv\xi+3w$, where $\gamma=0$ denotes the non-interacting limit, the GOLD sample gives $\gamma=1.453^{+1.297}_{-1.597}$ for the PantheonPlus-SH0ES and $\gamma=-0.634^{+1.668}_{-2.486}$ for the PantheonPlus-only. Although the posterior medians correspond to opposite directions of energy transfer, neither route excludes $\gamma=0$ at 68\% credibility, and the reconstructed interaction rate remains consistent with zero over the redshift range considered. Replacing the GOLD sample with the FULL sample produces negligible changes in the interaction constraints. Moreover, $w$CDM and CPL achieve likelihood improvements comparable to that of $\xi$IDE, while the information criteria do not consistently favor the interacting model. A redshift-bin diagnostic finds no significant redshift evolution of the Amati relation. We find no compelling evidence for a dark sector interaction that is robust to the choice of SN calibration or specifically favored over noninteracting dark energy extensions.
\footnotetext{Corresponding author.}}

\begin{document}
\maketitle
\flushbottom

\section{Introduction}\label{sec:introduction}

The discovery of the accelerated expansion of the Universe makes the physical nature of dark energy one of the central questions in modern cosmology \cite{Riess1998,Perlmutter1999}.  The spatially flat $\Lambda$CDM model provides an economical description of a wide range of observations \cite{Planck2018}. Nevertheless, long-standing theoretical problems, including the cosmological constant problem\cite{Weinberg1989} and the cosmic coincidence problem\cite{Chimento2003}, remain unresolved. The discrepancy between early- and late-Universe determinations of the Hubble constant further motivates tests of physics beyond the standard model\cite{Planck2018,Riess2022,Valentino2025}. Numerous extensions of $\Lambda$CDM have been proposed, including interacting dark energy (IDE). A simple extension of $\Lambda$CDM is $w$CDM, in which the equation of state of dark energy is constant but is allowed to differ from $-1$. The Chevallier–Polarski–Linder (CPL) parametrization\cite{Chevallier2001,Linder2003} generalizes this description to a time-varying dark energy. IDE models allow a nongravitational exchange of energy between dark matter and dark energy\cite{Amendola2000,Caldera2009,Wangbin2024}. Such an interaction modifies the evolution of the individual dark sector densities and the late-time expansion history, and has been investigated as a possible ways of alleviating the coincidence problem and cosmological tensions\cite{Costa2014,Pan2019,Valentino2021,Abdalla2022}.

Late-time observations provide complementary information about this background evolution. Type Ia supernovae (SNe Ia) constrain relative luminosity distances unless their absolute magnitude is calibrated. BAO measurements constrain distances relative to the sound horizon $r_d$ and do not determine $H_0$ and $r_d$ separately\cite{Heavens2014,Li2020}. Cosmic chronometers (CC) provide direct estimates of $H(z)$ from the differential ages of passively evolving galaxies\cite{Jimenez2002,Moresco2022}. Consequently, the inferred IDE constraints may depend on the treatment of the SN absolute calibration and the BAO ruler\cite{Benisty2024}. GRBs extend distance measurements beyond the redshift range of SNe Ia, but their empirical correlations must be calibrated before GRBs can be used as cosmological distance indicators\cite{Liang2008}. Wang and Liang\cite{WangLiang2024} constructed the GOLD and FULL samples of long GRBs from 15 years of \textit{Fermi}/GBM observations. The GOLD sample contains 123 long GRBs over $ 0.0785 \leq z \leq 5.6 $, and the FULL sample contains 151 events and extends to $z=8.2$. They calibrated the Amati relation at low redshift using distances reconstructed from CC data and combined the resulting high-redshift GRB Hubble diagram with PantheonPlus SNe Ia to constrain cosmological models.

Several recent studies have used late-time observations to constrain phenomenological IDE models. Nong and Liang\cite{Nong2024} constrained the IDE model using a calibrated GRB Hubble diagram together with PantheonPlus SNe Ia. Benisty et al.\cite{Benisty2024} investigated a linear interaction using calibrated and uncalibrated SN samples while analytically marginalizing the late-time scale parameters, demonstrating that the inferred coupling can depend on the treatment of the absolute calibration. Zhu et al.\cite{Zhu2026} constrained the same $\xi$IDE model using high-redshift \textit{Fermi} GRB Hubble diagram and DESI DR2 BAO, with an external Planck prior on $r_d$. Figueruelo et al.\cite{Figueruelo2026} examined five linear and three nonlinear IDE kernels using PantheonPlus SNe Ia, DESI DR2 BAO, cosmic chronometers, and BBN information. Meng et al.\cite{Meng2026} combined \textit{Fermi} GRBs with PantheonPlus SNe Ia, propagated the full covariance associated with the Amati-relation calibration, and tested two linear interaction models.

Adapting the phenomenological ansatz introduced by Dalal et al.\cite{Dalal2001}, we consider an interacting dark sector model in which the ratio of the dark-energy density to the dark-matter density evolves as $a^\xi$, while baryons remain separately conserved. The measurements of the observed GRB spectral peak energy and bolometric fluence are included directly in the likelihood\cite{Khadka2020}. The Amati intercept, slope, and intrinsic scatter are sampled jointly with the cosmological parameters\cite{Cao2024}. This joint treatment accounts for the mutual dependence between the Amati-relation and cosmological parameters and avoids adopting a cosmological model in a separate calibration step\cite{Amati2008,Khadka2021}. We combine the GRB measurements with PantheonPlus SNe Ia~\cite{Scolnic2022,Brout2022}, DESI DR2 BAO~\cite{DESI2025}, and an updated CC compilation. Two complementary SN routes are considered. The PantheonPlus-SH0ES route retains the SN absolute calibration, whereas in the PantheonPlus-only route, the SN absolute-magnitude parameter is analytically marginalized. The BAO ruler is left free. we sample $r_d$ in the PantheonPlus-SH0ES and $hr_d$ in the PantheonPlus-only. We extend the calibration-sensitivity question studied by Benisty et al.\cite{Benisty2024} for a linear IDE model to the specific nonlinear $\xi$IDE parametrization considered here. A redshift-bin diagnostic is applied to test whether the fitted Amati slope shows evidence of redshift evolution.

This paper is organized as follows. Section~\ref{sec:models} introduces the cosmological models. Section~\ref{sec:data-method} describes the datasets and methodology, including the two SN calibration, likelihood construction, and the Amati redshift-evolution diagnostic. Section~\ref{sec:results} presents the results of Amati redshift-evolution diagnostic and $\xi$IDE constraints. Section~\ref{sec:sum_con} summarizes the main results and conclusions. Appendix ~\ref{app:h0-rd} presents the joint constraints on $H_0$ and $r_d$ for the two SN calibration.

\section{Cosmological models}\label{sec:models}

We assume a homogeneous and isotropic, spatially flat Friedmann--Lema\^{\i}tre--Robertson--Walker universe. The dimensionless expansion rate is
\begin{equation}
    E(z) \equiv \frac{H(z)}{H_0},
\end{equation}
where $H_0$ is the Hubble constant. The present-day density parameters satisfy
\begin{equation}
    \Omega_{m0}+\Omega_{{\rm de},0}=1,
    \qquad
    \Omega_{m0}=\Omega_{b0}+\Omega_{c0},
\end{equation}
where $\Omega_{m0}$, $\Omega_{b0}$, $\Omega_{c0}$ and $\Omega_{{\rm de},0}$ denote the density parameters of total matter, baryons, (cold) dark matter, and dark energy, respectively.

\subsection{Non-interacting models}

For the $\Lambda$CDM model, the background expansion is
\begin{equation}
    E^2(z) = 1 - \Omega_{m0} + \Omega_{m0}(1+z)^3.
\end{equation}
For $w$CDM,
\begin{equation}
    E^2(z) = \Omega_{m0}(1+z)^3 + (1-\Omega_{m0})(1+z)^{3(1+w)},
\end{equation}
where $w$ is the constant dark energy equation-of-state. Dynamical dark energy is modeled using the Chevallier–Polarski–Linder (CPL) parametrization\cite{Chevallier2001,Linder2003},
\begin{equation}
    w(z) = w_0 + \frac{z}{1+z} w_a,
\end{equation}
which gives
\begin{equation}
    E^2(z) = \Omega_{m0}(1+z)^3 + (1-\Omega_{m0})(1+z)^{3(1+w_0+w_a)}\exp\left[-\frac{3w_a z}{1+z}\right],
\end{equation}
where $w_0$ is the present-day value of the dark energy equation-of-state parameter, while $w_a$ controls its redshift evolution.

\subsection{Phenomenological IDE model}

In the phenomenological IDE model considered here, energy exchange between dark matter and dark energy is described by a coupling term $Q$,
\begin{align}
    \dot{\rho}_c + 3H\rho_c &= Q, \\
    \dot{\rho}_{\mathrm{de}} + 3H(1+w)\rho_{\mathrm{de}} &= -Q ,
\end{align}
where $\rho_c$ and $\rho_{\mathrm{de}}$ are the energy densities of dark matter and dark energy. With this sign convention, $Q>0$ denotes energy transfer from dark energy to dark matter. Baryons are separately conserved.

We consider a $\xi$IDE model\cite{Dalal2001,Zimdahl2006} and assume that the dark sector density ratio evolves as
\begin{equation}
    \frac{\rho_{\mathrm{de}}(z)}{\rho_c(z)} = r_0(1+z)^{-\xi}, 
    \qquad
    r_0 = \frac{\Omega_{\mathrm{de},0}}{\Omega_{c0}} .
\end{equation}

For a constant $w$, this ansatz implies\cite{Chimento2010,Westhuizen2025,Westhuizen2025b}
\begin{equation}
    Q = -H(\xi + 3w) \frac{\rho_c \rho_{\mathrm{de}}}{\rho_c + \rho_{\mathrm{de}}} .
\end{equation}

The background expansion corresponding to $\xi$IDE is
\begin{equation}\label{eq:xiide-e2}
    E^2(z)=\Omega_{b0}(1+z)^3+(1-\Omega_{b0})(1+z)^3 \left[\frac{1+r_0(1+z)^{-\xi}}{1+r_0}\right]^{-3w/\xi}.
\end{equation}

It is convenient to define the interaction combination
\begin{equation}
    \gamma\equiv\xi+3w,
\end{equation}
when $\gamma=0$, the interaction vanishes and Eq.\eqref{eq:xiide-e2} reduces to the $w$CDM. $\Lambda$CDM corresponds to $(w,\xi,\gamma)=(-1,3,0)$. We sample $(w,\gamma)$ and derive $\xi=\gamma-3w$.

To characterize the coupling by a dimensionless quantity that can be compared across redshifts, we define the dark sector fractions and normalized interaction rate
\begin{equation}\label{eq:interaction-rate}
    f_c(z)=\frac{\rho_c}{\rho_c+\rho_{\rm de}},
    \qquad
    f_{\rm de}(z)=1-f_c(z),
    \qquad
    \mathcal{I}(z)\equiv\frac{Q}{H(\rho_c+\rho_{\rm de})} =-\gamma f_c(z)f_{\rm de}(z).
\end{equation}
With our sign convention, $\mathcal{I}>0$ denotes energy transfer from dark energy to dark matter and $\mathcal{I}<0$ denotes the reverse direction. At the present epoch,
\begin{equation}\label{eq:interaction-rate-zero}
    \mathcal{I}_0=-\gamma
    \frac{\Omega_{c0}\Omega_{{\rm de},0}}{(\Omega_{c0}+\Omega_{{\rm de},0})^2}.
\end{equation}

\section{Data and methodology}\label{sec:data-method}
\subsection{Fermi long GRBs}

We use the GOLD and FULL samples compiled from 15 yr of \textit{Fermi}/GBM observations by Wang and Liang \cite{WangLiang2024}. The GOLD sample contains 123 long GRBs over $ 0.0785 \leq z \leq 5.6 $, and the FULL sample contains 151 events and extends to $z=8.2$. Previous analyses have shown that the two samples yield broadly consistent cosmological constraints \cite{WangLiang2024,Zhu2026,Meng2026}. Because GOLD is the more restrictive subset, we adopt it for the fiducial analysis and use FULL only to assess sensitivity to the GRB selection.

The Amati relation connects the rest-frame peak energy $E_{{\rm p},i}$ and the isotropic energy $E_{{\rm iso},i}$ \cite{Amati2002}:
\begin{equation}
    y_i = a + b x_i ,
    \qquad
    x_i = \log_{10}\left(\frac{E_{p,i}}{300\,\mathrm{keV}}\right),
    \qquad
    y_i = \log_{10}\left(\frac{E_{\mathrm{iso},i}}{\mathrm{erg}}\right).
\end{equation}

The $E_{p,i}$ and $E_{\mathrm{iso},i}$ are given by
\begin{equation}
    E_{p,i}=E^{\rm obs}_{p,i}(1+z_i),
    \qquad
    E_{{\rm iso},i}=\frac{4\pi D_L^2(z_i)S_{{\rm bolo},i}}{1+z_i},
\end{equation}
where $E^{\mathrm{obs}}_{p,i}$ and $S_{\mathrm{bolo},i}$ are the measured peak energy and bolometric fluence, respectively. For each event, the likelihood is evaluated from the observables $(z_i,E_{p,i},S_{{\rm bolo},i})$ and their uncertainties. Defining $\Delta_i=y_i-a-bx_i$, the total variance is
\begin{equation}
    s_i^2 = \left(\frac{\sigma_{S_{\mathrm{bolo},i}}}{S_{\mathrm{bolo},i}\ln 10} \right)^2 + b^2 \left( \frac{\sigma_{E_{p,i}}} {E_{p,i}\ln 10} \right)^2 + \sigma_{\mathrm{int}}^2 ,
\end{equation}
where $\sigma_{\rm int}$ denotes the intrinsic scatter of the Amati relation, accounting for dispersion that is not captured by the measurement uncertainties in $ E_{{\rm p},i}$ and $S_{{\rm bolo},i}$.

The GRB likelihood is given by\cite{Agostini2005}
\begin{equation}\label{eq:grb-like}
    -2\ln \mathcal{L}_{\mathrm{GRB}} = \sum_i \left( \frac{\Delta_i^2}{s_i^2} + \ln s_i^2 \right).
\end{equation}

The parameters $(a,b,\sigma_{\mathrm{int}})$ are sampled jointly with the cosmological parameters.

\subsection{PantheonPlus SNe Ia}

We use the PantheonPlus data release \cite{Scolnic2022,Brout2022} in two complementary forms. The purpose of this split is to determine whether the inferred cosmological constraints depend on the SN absolute calibration.

In the PantheonPlus-SH0ES route, we use the full PantheonPlus SH0ES data of 1701 SN light-curve entries and the full covariance matrix\footnote{\url{https://github.com/PantheonPlusSH0ES/DataRelease/tree/main/Pantheon+_Data/4_DISTANCES_AND_COVAR}}, which includes both statistical and systematic uncertainties. The SN contribution to the total $\chi^2$ is

\begin{equation}
    \chi^2_{\rm SN}
    =\Delta\boldsymbol{\mu}^{\rm T}C_{\rm SN}^{-1}\Delta\boldsymbol{\mu},
    \qquad
    \Delta\boldsymbol{\mu}=\boldsymbol{\mu}_{\rm SH0ES}-\boldsymbol{\mu}_{\rm th}.
\end{equation}

In the PantheonPlus-only route, we follow the PantheonPlus cosmology construction\footnote{\url{https://github.com/PantheonPlusSH0ES/DataRelease/tree/main/Pantheon\%2B_Data/5_COSMOLOGY/cosmosis_likelihoods}}. We retain entries with $z_{\rm HD}>0.01$, crop the full covariance matrix with the same mask, and obtain 1590 light-curve entries. The corrected apparent magnitude satisfies
\begin{equation}
    m_{b,{\rm corr}}=\mu_{\rm th}+M,
\end{equation}
where $M$ is the nuisance SN absolute-magnitude parameter. We remove $M$ from our fits by analytic marginalization, and derive the $\chi^2_{\rm SN,M}$ \cite{Goliath2001,Pietro2003,Lazkoz2005,Conley2011,Benisty2024},
\begin{equation}\label{eq:sn-marg}
    \chi^2_{\rm SN,M} = A - \frac{B^2}{E},
\end{equation}
where
\begin{equation}
    A=\Delta\boldsymbol{m}^{\rm T}C_{\rm SN}^{-1}\Delta\boldsymbol{m},
    \qquad
    B=\Delta\boldsymbol{m}^{\rm T}C_{\rm SN}^{-1}\boldsymbol{1},
    \qquad
    E=\boldsymbol{1}^{\rm T}C_{\rm SN}^{-1}\boldsymbol{1}.
\end{equation}
and $\Delta\boldsymbol{m}=\boldsymbol{m}_{b,{\rm corr}}-\boldsymbol{\mu}_{\rm th}$. This constrains the shape of the SN distance--redshift relation but supplies no independent absolute distance scale.

\subsection{DESI DR2 BAO}

We use the full DESI DR2 BAO dataset\cite{DESI2025}, comprising 13 measurements of $D_M(z)/r_d$, $D_H(z)/r_d$, and $D_V(z)/r_d$ with their released covariance matrix\footnote{\url{https://github.com/CobayaSampler/bao_data/tree/master/desi_bao_dr2}}. For a spatially flat universe,
\begin{align}
    D_M(z)&=\frac{c}{H_0}\int_0^z\frac{dz'}{E(z')},\\
    D_H(z)&=\frac{c}{H_0E(z)},\\
    D_V(z)&=\left[zD_M^2(z)D_H(z)\right]^{1/3}.
\end{align}

The $\chi^2$ of BAO is
\begin{equation}
    \chi^2_{\rm BAO}=(\boldsymbol{D}_{\rm obs}-\boldsymbol{D}_{\rm th})^{\rm T}C_{\rm BAO}^{-1} (\boldsymbol{D}_{\rm obs}-\boldsymbol{D}_{\rm th}).
\end{equation}

For each BAO distance variable \(D_X\in\{D_M,D_H,D_V\}\), the theoretical quantity can be written as
\begin{equation}
    \frac{D_X(z)}{r_d} = \frac{c}{H_0 r_d}\,\mathcal{F}_X[E;z],
\end{equation}
where $\mathcal{F}_X$ depends on the dimensionless expansion history\cite{Benisty2024}. The BAO data constrain the combined scale $H_0 r_d$, or equivalently $h r_d$, much more directly than $H_0$ and $r_d$ separately. Related low-redshift analyses have likewise inferred $H_0$ and $r_d$ jointly without fixing $r_d$ to its CMB-derived value\cite{Arendse2020,Nagpal2025}. For the PantheonPlus-SH0ES, the SN absolute calibration and the CC data provide absolute-scale information. For the pantheonPlus-only, analytic marginalization over the SN absolute-magnitude moves the SN absolute scale information, and the CC measurements provide the principal absolute $H(z)$ information. We sample $r_d$ in the PantheonPlus-SH0ES route. In the PantheonPlus-only route, we sample
\begin{equation}
    h r_{d}\equiv \frac{H_0}{100\,{\rm km\,s^{-1}\,Mpc^{-1}}} r_d,
\end{equation}
and derive $r_d$ from
\begin{equation}
    r_d = \frac{h r_d}{h}.
\end{equation}

\subsection{Cosmic chronometers}

We use an updated and curated CC compilation containing 32 $H(z)$ measurements over $ 0.07 \leq z \leq 1.965 $, as listed in Table~\ref{tab:cc_updated}. The construction of this CC compilation is guided by the updated sample-selection discussions\cite{Moresco2024,Jiao2025}. We begin with the 26-point curated selection of Jiao et al.\cite{Jiao2025}, which removes the original Simon et al. determinations, avoids the simultaneous use of measurements based on overlapping galaxy samples, and replaces the commonly quoted extrapolated value $69\pm12$ at $z=0.09$ by the value $H(z=0.09)=72\pm13\,{\rm km\,s^{-1}\,Mpc^{-1}}$. Twenty-four of these measurements correspond to entries in the LUDB release\footnote{\url{https://github.com/MauLoHdz/LUDB/tree/main/Data/CC}} \cite{Lopez2025}, the measurements at $z=0.50$\cite{Loubser2025} and $z=0.75$\cite{Jimenez2023} are added separately. We further include six recent measurements\cite{Wang2026,Loubser2025,Tomasetti2026,Pradhan2026}.

We construct the covariance of the updated CC compilation from the $33\times33$ covariance $C_{\rm LUDB}$ distributed with the LUDB release. For the 24 retained LUDB measurements, we extract the submatrix
\begin{equation}
    C_{\rm base}=C_{\rm LUDB}[I, I].
\end{equation}
where $I$ denotes the LUDB row numbers,
\begin{equation}
I=\{1,2,3,5,6,7,9,10,11,13,14,15,16,17,18,19,20,21,23,24,26,27,29,33\}.
\end{equation}

The remaining eight measurements, which are the final eight rows of Table~\ref{tab:cc_updated}, are represented by the diagonal block $C_{\rm add}$. The $32\times32$ covariance matrix of the updated CC compilation is
\begin{equation}\label{eq:cc-block-cov}
    C_{\rm CC}=
    \begin{pmatrix}
        C_{\rm base} & 0\\
        0 & C_{\rm add}
    \end{pmatrix}.
\end{equation}

The $\chi^2$ of CC is written as
\begin{equation}
    \chi^2_{\rm CC} = \Delta \mathbf{H}^{T} C_{\rm CC}^{-1} \Delta \mathbf{H},
\end{equation}
where $\Delta H = H_{\rm obs} - H_0 E(z)$, and $C_{\rm CC}$ is the covariance matrix of the CC compilation.

\small
\setlength{\LTcapwidth}{\textwidth}
\begin{longtable}{@{}ccccl@{}}
\caption{Updated 32-point CC compilation used in this work. $H(z)$ and $\sigma_H$ are in ${\rm km\,s^{-1}\,Mpc^{-1}}$. The data span $z=0.07-1.965$ using various differential age methods: full spectral fitting (F), D4000 (D), and Machine Learning (ML). A dagger marks the eight measurements represented by the diagonal block $C_{\rm add}$ in Eq.~\eqref{eq:cc-block-cov}. The asterisk marks the revised $z=0.09$ data.}\label{tab:cc_updated}\\
\toprule
$z$ & $H(z)$ & $\sigma_H$ & Method & Reference \\
\midrule
\endfirsthead
\multicolumn{5}{c}{\tablename~\thetable\ (continued)}\\
\toprule
$z$ & $H(z)$ & $\sigma_H$ & Method & Reference \\
\midrule
\endhead
\midrule
\multicolumn{5}{r}{Continued on next page}\\
\endfoot
\bottomrule
\endlastfoot
0.07 & 69.0   & 19.6  & F  & Zhang et al. \cite{Zhang2014} \\
0.09$^{*}$ & 72   & 13  & F  & Jimenez et al. \cite{Jimenez2003} \\
0.12 & 68.6   & 26.2  & F  & Zhang et al. \cite{Zhang2014} \\
0.1791 & 75   & 4   & D  & Moresco et al. \cite{Moresco2012} \\
0.1993 & 75   & 5   & D  & Moresco et al. \cite{Moresco2012} \\
0.20 & 72.9   & 29.6  & F  & Zhang et al. \cite{Zhang2014} \\
0.28 & 88.8   & 36.6  & F  & Zhang et al. \cite{Zhang2014} \\
0.3519 & 83   & 14  & D  & Moresco et al. \cite{Moresco2012} \\
0.3802 & 83.0   & 13.5  & D  & Moresco et al. \cite{Moresco2016} \\
0.4004 & 77.0   & 10.2  & D  & Moresco et al. \cite{Moresco2016} \\
0.4247 & 87.1   & 11.2  & D  & Moresco et al. \cite{Moresco2016} \\
0.4497 & 92.8   & 12.9  & D  & Moresco et al. \cite{Moresco2016} \\
0.47 & 89   & 49.6  & F  & Ratsimbazafy et al. \cite{Ratsimbazafy2017} \\
0.4783 & 80.9   & 9   & D  & Moresco et al. \cite{Moresco2016} \\
0.48 & 97   & 62  & F  & Stern et al. \cite{Stern2010} \\
0.5929 & 104  & 13  & D  & Moresco et al. \cite{Moresco2012} \\
0.6797 & 92   & 8   & D  & Moresco et al. \cite{Moresco2012} \\
0.7812 & 105  & 12  & D  & Moresco et al. \cite{Moresco2012} \\
0.8754 & 125  & 17  & D  & Moresco et al. \cite{Moresco2012} \\
0.88 & 90   & 40  & F  & Stern et al. \cite{Stern2010} \\
1.037 & 154  & 20  & D  & Moresco et al. \cite{Moresco2012} \\
1.26 & 135  & 65  & F  & Tomasetti et al. \cite{Tomasetti2023} \\
1.363 & 160  & 33.6  & D  & Moresco \cite{Moresco2015} \\
1.965 & 186.5  & 50.4  & D  & Moresco \cite{Moresco2015} \\
0.120$^{\dagger}$ & 71.33  & 4.20  & F  & Wang et al. \cite{Wang2026} \\
0.460$^{\dagger}$ & 88.48  & 12.33 & D  & Loubser et al. \cite{Loubser2025} \\
0.5$^{\dagger}$ & 72.1   & 34.7  & D  & Loubser et al. \cite{Loubser2025} \\
0.542$^{\dagger}$ & 66.00 & ${}_{-32}^{+82}$ & F
& Tomasetti et al. \cite{Tomasetti2026} \\
0.650$^{\dagger}$ & 93.68  & 30.22 & D+Bayes  & Pradhan et al. \cite{Pradhan2026} \\
0.670$^{\dagger}$ & 119.45 & 17.82 & D  & Loubser et al.\cite{Loubser2025} \\
0.75$^{\dagger}$ & 105.0  & 10.8  & ML & Jimenez et al. \cite{Jimenez2023} \\
0.830$^{\dagger}$ & 108.28 & 18.13 & D  & Loubser et al.
\cite{Loubser2025} \\
\end{longtable}
\normalsize

\subsection{Baryons and the joint likelihood}

The late-time data used here do not independently separate baryons from dark matter with high precision. We adopt the Planck Gaussian prior on $\omega_b\equiv\Omega_b h^2=0.02237\pm0.00015$\cite{Planck2018} to constrain the baryonic contribution to the late-time universe, motivated by previous analyses that replacing this prior with a broad uniform prior left the principal cosmological constraints essentially unchanged\cite{Meng2026}. Similar stability under reasonable variations of the baryon-density prior has also been reported in other low-redshift cosmological analyses\cite{Zhang2019,Lin2017,Adame2025}.

We compute
\begin{equation}
    \Omega_{b0}=\frac{\omega_b}{h^2},
    \qquad
    h=\frac{H_0}{100\,{\rm km\,s^{-1}\,Mpc^{-1}}}.
\end{equation}

The total likelihood for the PantheonPlus-SH0ES is
\begin{equation}
    -2\ln\mathcal{L}_{\rm total}=-2\ln\mathcal{L}_{\rm GRB}
    +\chi^2_{\rm SN}+\chi^2_{\rm BAO}+\chi^2_{\rm CC}.
\end{equation}
For the PantheonPlus-only, $\chi^2_{\rm SN}$ is replaced by the marginalized expression in Eq.~\eqref{eq:sn-marg}.

\subsection{Sampling and model comparison}

We employ the MCMC analysis in the \texttt{emcee} package\cite{Foreman2013}. The cosmological parameters are sampled jointly with $(a,b,\sigma_{\rm ext})$ and the appropriate BAO scale parameter, $r_d$ or $hr_d$. For $\xi$IDE we sample $\gamma=\xi+3w$ and derive $\xi$. During sampling, we impose physicality conditions requiring positive dark sector densities and $E^2(z)>0$ over the redshift range used in the data.

For models fitted to the same data combination, we report the Akaike and Bayesian information criteria,
\begin{equation}
    {\rm AIC}=-2\ln\mathcal{L}_{\max}+2p,
    \qquad
    {\rm BIC}=-2\ln\mathcal{L}_{\max}+p\ln N,
\end{equation}
where $p$ is the number of fitted parameters and $N$ is the number of data points. The Gaussian prior on $\omega_b$ is used in posterior sampling but is excluded from the data likelihood entering AIC/BIC. We quote
\begin{equation}
    \Delta{\rm AIC}={\rm AIC}_{\rm model}-{\rm AIC}_{\Lambda{\rm CDM}},
    \qquad
    \Delta{\rm BIC}={\rm BIC}_{\rm model}-{\rm BIC}_{\Lambda{\rm CDM}}.
\end{equation}

\subsection{Amati redshift-evolution diagnostic}

In this work, we assume one redshift-independent Amati relation. We also test for possible redshift evolution with a separate bin-wise diagnostic\cite{Ghirlanda2008,Wang2011,Dai2021,Han2024}. More recent studies have continued to test the redshift dependence of GRB correlations by refitting the Amati parameters in separate redshift subsamples\cite{Singh2025,Hasan2026}. We adopt flat $\Lambda$CDM as the representative cosmological model, using the GOLD GRB sample, the PantheonPlus-SH0ES, the DESI DR2 BAO, and the updated CC compilation. Using the same fixed set of parameters, we recalculate $E_{\rm iso}$ separately for the GOLD and FULL samples.

The GRBs in each sample are divided into six redshift bins: [0,0.5), [0.5,1.0), [1.0,1.5), [1.5,2.0), [2.0,3.0), and [3.0,9.0]. Within each bin, we refit $(a_i,b_i,\sigma_{{\rm int},i})$ with the same errors-in-variables likelihood as in Eq.~\eqref{eq:grb-like}. We associate each fit with the median redshift of its bin and fit $b_i=b_0+b_1z_i$. The primary evolution statistic is $b_1=db/dz$. We also record the trends $a_1=da/dz$ and $s_1=d\sigma_{\rm int}/dz$ as ancillary diagnostics. We use a nonparametric case bootstrap to estimate the uncertainties in the fitted parameters and their redshift trends\cite{Efron1979,Freedman1981,Feigelson1992,Andrae2010}. For each sample, we generate 2000 bootstrap resamples within each fixed redshift interval and refit the Amati relation to every resampled data set. For each bootstrap realization, the resulting bin-wise Amati slopes are combined in the final fit of $b(z)$, thereby constructing the bootstrap distribution of $db/dz$. The same procedure is applied to $da/dz$ and $d\sigma_{\rm int}/dz$. The quoted confidence intervals are obtained from the corresponding bootstrap distributions\cite{DiCiccio1996}.

\section{Results}\label{sec:results}

\subsection{Test of Amati-slope redshift evolution}

\begin{figure}[h]
    \centering
    \includegraphics[width=0.92\textwidth]{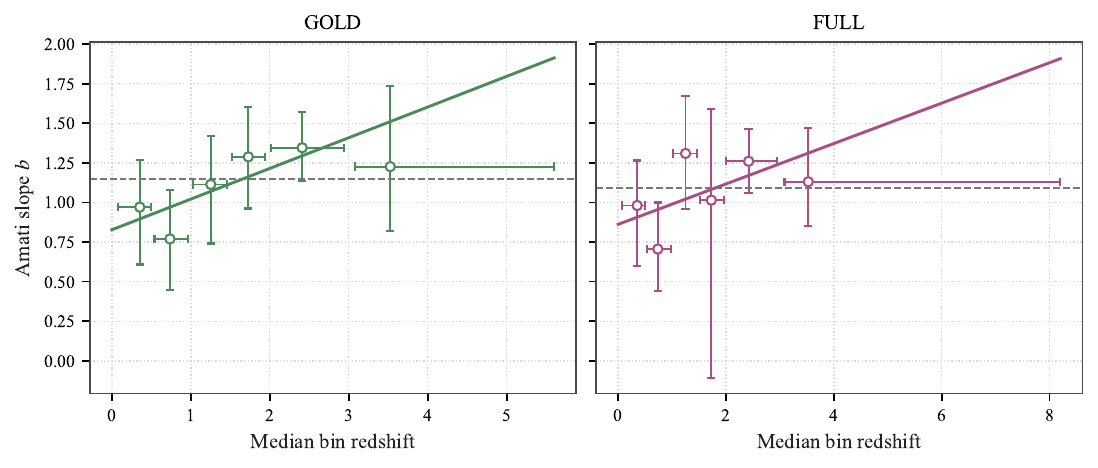}
    \caption{Bin-wise Amati-slope constraints for the GOLD and FULL samples under the fixed representative $\Lambda$CDM model. Points show the median of the Amati slope $b$ in each redshift interval. Vertical error bars denote the 68\% intervals, and horizontal bars show the actual redshift range of the GRBs in each interval. Solid lines show the linear trends corresponding to the median values of $db/dz$. Dashed lines show the best-fitting redshift-independent value of the Amati slope $b$, with more precisely determined redshift bins given greater weight. The 95\% intervals include $db/dz=0$ for both samples.}
    \label{fig:fig1}
\end{figure}

For the flat $\Lambda$CDM model, we obtain the best-fitting values $H_0=73.395~{\rm km\,s^{-1}\,Mpc^{-1}}$ and $\Omega_{\rm m0}=0.3094$. These values are held fixed when $E_{\rm iso}$ is recalculated separately for the GOLD and FULL samples. Table~\ref{tab:table2} and Figure~\ref{fig:fig1} shows the bin-wise estimates of the Amati slope obtained under the fixed representative $\Lambda$CDM model. The primary result for GOLD is
\begin{equation}
    \frac{db}{dz}=0.194^{+0.138}_{-0.132}.
\end{equation}

For FULL we obtain
\begin{equation}
    \frac{db}{dz}=0.128^{+0.116}_{-0.112}.
\end{equation}

Zero lies inside both 95\% confidence intervals. We find no statistically significant redshift evolution of the Amati slope.

\begin{table}[h]
\centering
\caption{Redshift-evolution diagnostics for the bin-wise Amati parameters under the fixed representative $\Lambda$CDM model. The quoted uncertainties and intervals are obtained from the bootstrap distributions.}
\label{tab:table2}
\begin{tabular}{clcc}
\hline
Samples & Parameters & Trend (68\% interval) & Trend (95\% interval) \\
\hline
\multirow{3}{*}{GOLD}
 & $da/dz$
 & $0.191^{+0.065}_{-0.067}$ & $[0.050,0.318]$ \\
 & $db/dz$
 & $0.194^{+0.138}_{-0.132}$ & $[-0.061,0.507]$ \\
 & $d\sigma_{\rm int}/dz$
 & $-0.074^{+0.026}_{-0.028}$ & $[-0.133,-0.023]$ \\
\hline
\multirow{3}{*}{FULL}
 & $da/dz$
 & $0.203^{+0.058}_{-0.060}$ & $[0.084,0.319]$ \\
 & $db/dz$
 & $0.128^{+0.116}_{-0.112}$ & $[-0.084,0.379]$ \\
 & $d\sigma_{\rm int}/dz$
 & $-0.070^{+0.021}_{-0.024}$ & $[-0.117,-0.027]$ \\
\hline
\end{tabular}
\end{table}

The ancillary diagnostics reveal additional redshift structure in the observed samples. Although no statistically significant trend is found in the Amati slope $b$, the fitted $a$ increases and $\sigma_{\rm int}$ decreases across the redshift bins. Since the present analysis does not explicitly model residual selection effects associated with GRB detection, spectral analysis, and redshift measurement, etc, the trends in $a$ and $\sigma_{\rm int}$ cannot interpreted as evidence for intrinsic redshift evolution of the Amati relation\cite{WangLiang2024,Ghirlanda2008,Heussaff2013,Singh2025}.

\subsection{Constraints on the $\xi$IDE model}

\begin{figure}[h]
    \centering
    \includegraphics[width=0.97\textwidth]{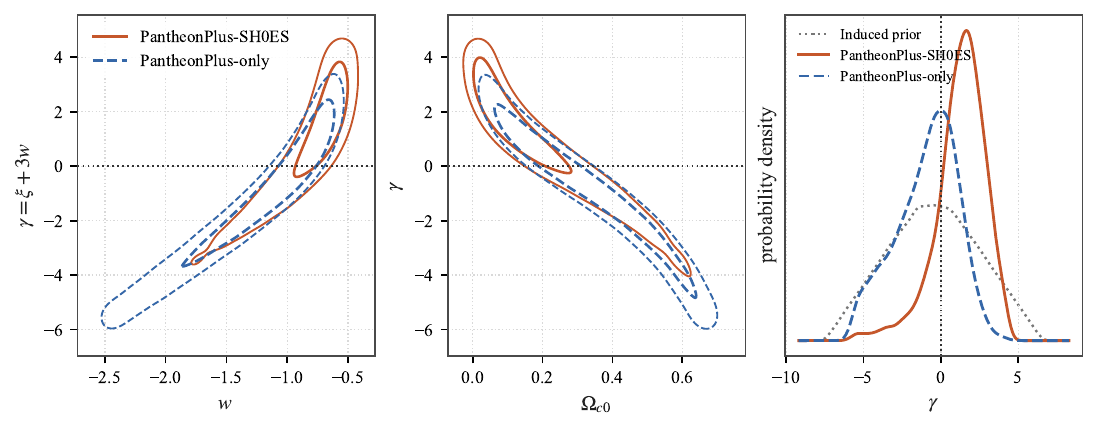}
    \caption{Constraints on the $\xi$IDE parameters from the GOLD sample. The first two panels show 68\% and 95\% credible contours in the $(w,\gamma)$ and $(\Omega_{c0},\gamma)$ planes. The third panel compares the $\gamma$ posteriors with the prior induced by the uniform priors in $(w,\xi)$. Dotted reference lines mark the non-interacting value $\gamma=0$.}
    \label{fig:fig2}
\end{figure}

We imposed independent uniform bounds on $w$ and $\xi$, under the transformation $\gamma=\xi+3w$, the rectangular region ($w$, $\xi$)  maps to a slanted, non-rectangular region in the sampled ($w$, $\gamma$) plane and induces a non-uniform prior for $\gamma$. Monte Carlo samples from this induced prior give a 95\% interval $-5.943<\gamma<5.040$ and $P_{\rm prior}(\gamma>0)=0.443$. The gray curve in the right panel of Figure~\ref{fig:fig2} shows this induced prior, which provides a reference for assessing how much the posterior constraint on \(\gamma\) is driven by the data rather than by the transformed parameter support.

Table~\ref{tab:table3} presents the main $\xi$IDE constraints from the GOLD sample. For both the PantheonPlus-SH0ES and PantheonPlus-only, the non-interacting value $\gamma=0$ lies within the 68\% credible interval. The corresponding posterior sign probabilities $P(\gamma>0)$ are $0.822$ and $0.367$, indicate opposite posterior tendencies but do not constitute statistically significant evidence for a nonzero interaction.

\begin{table}[h]
\centering
\caption{Constraints on the main $\xi$IDE parameters (68\% credible intervals) obtained with the GOLD sample. The 95\% interval of $\gamma$ is listed separately.}
\label{tab:table3}
\begin{tabular}{lcc}
\toprule
Parameters
& PantheonPlus-SH0ES
& PantheonPlus-only \\
\midrule
$\Omega_{\mathrm{c}0}$
& $0.111^{+0.162}_{-0.067}$
& $0.321^{+0.227}_{-0.177}$ \\

$w$
& $-0.681^{+0.085}_{-0.214}$
& $-1.041^{+0.278}_{-0.658}$ \\

$\gamma$
& $1.453^{+1.297}_{-1.597}$
& $-0.634^{+1.668}_{-2.486}$ \\

$\xi$
& $3.493^{+1.059}_{-0.968}$
& $2.493^{+0.845}_{-0.519}$ \\

\midrule
$\gamma$ (95\% interval)
& $[-2.999,\;3.835]$
& $[-5.165,\;2.374]$ \\

$P(\gamma>0)$
& $0.822$
& $0.367$ \\

Median $\mathcal{I}_0$
& $-0.120$
& $0.142$ \\

$P(Q_0>0)$
& $0.178$
& $0.633$ \\
\bottomrule
\end{tabular}
\end{table}

The parameter $\gamma$ determines the direction of energy transfer, but its sign probability does not describe the strength of the interaction or how that strength changes with redshift. We define the dimensionless interaction rate

\begin{equation}
    \mathcal I(z) = \frac{Q}{H(\rho_{\rm c}+\rho_{\rm de})}
    = -\gamma
    \frac{\Omega_c \Omega_{\rm de}}
    {(\Omega_c+\Omega_{\rm de})^2}.
\end{equation}

Although $\gamma$ is constant in this model, $\mathcal I(z)$ varies with redshift because the fractional contributions of dark sectors evolve. Under our sign convention, $\mathcal I>0$ corresponds to energy transfer from dark energy to dark matter and $\mathcal I<0$ denotes the reverse direction. For any individual model realization, the direction does not change with redshift, only the magnitude of the interaction evolves.

\begin{figure}[h]
    \centering
    \includegraphics[width=0.97\textwidth]{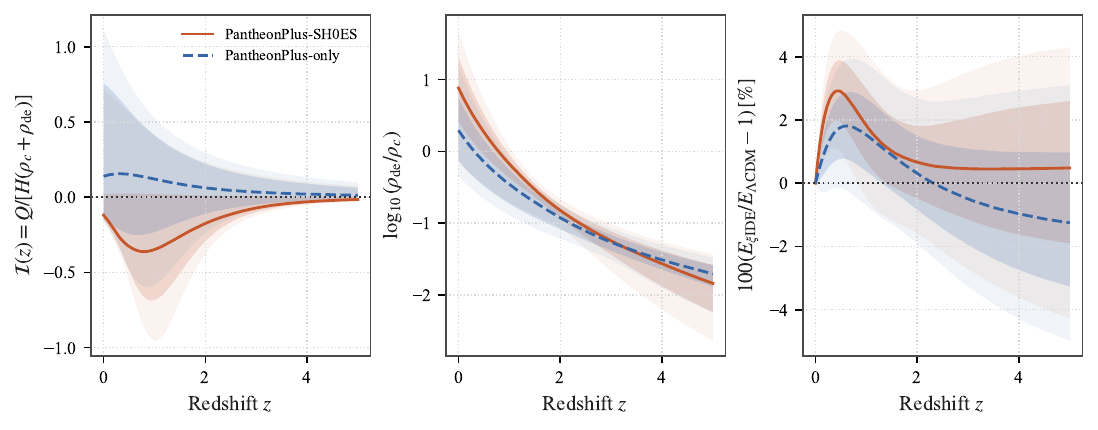}
    \caption{Reconstruction of the $\xi$IDE background. The left panel shows the dimensionless interaction rate $\mathcal I(z)=Q/[H(\rho_{\rm c}+\rho_{\rm de})]$, positive values correspond to energy transfer from dark energy to dark matter. The middle panel shows the evolve of dark sector density ratio $\rho_{\rm de}/\rho_{\rm c}$. The right panel shows the difference between the $\xi$IDE prediction for $E(z)$ and the $\Lambda$CDM prediction. The solid curve uses the PantheonPlus-SH0ES, and the dashed curve uses the PantheonPlus-only. The dark and light bands show the 68\% and 95\% credible regions, respectively. Horizontal dotted lines mark zero interaction and zero expansion-rate difference.}
    \label{fig:fig3}
\end{figure}

Figure~\ref{fig:fig3} illustrates this evolution. The left panel shows $\mathcal I(z)$ and measures the interaction strength relative to the total dark sector density. The middle panel shows the density ratio $\rho_{\rm de}(z)/\rho_{\rm c}(z)$, which describes how the relative abundance of the two dark sector components changes with redshift. For each SN route, the right panel shows the percentage difference between the $\xi$IDE prediction for $E(z)$ and the route-matched $\Lambda$CDM reference, 
\begin{equation}
    \Delta_E(z)\equiv100\left[\frac{E_{\xi{\rm IDE}}(z)} {E_{\Lambda{\rm CDM}}(z)} -1 \right] \%,
\end{equation}
where ${E_{\Lambda{\rm CDM}}(z)}$ denotes the posterior-median $\Lambda$CDM expansion history inferred from the same SN calibration and data. This inference is held fixed at its posterior median, so its posterior uncertainty is not propagated into the displayed bands. For the PantheonPlus-SH0ES, the median interaction rate changes from $\mathcal I(0)=-0.120$ at the present epoch to $\mathcal I(1)=-0.349$ at $z=1$. However, the 95\% credible interval at $z=1$, $-0.952 < \mathcal I(1) < 0.413$, includes the non-interacting value $\mathcal I=0$. For the PantheonPlus-only, the median values are $\mathcal I(0)=0.142$ and $\mathcal I(1)=0.121$, and their positive sign indicates a tendency toward energy transfer in the opposite direction. The 95\% credible intervals are $-0.157 < \mathcal I(0) < 1.110$ and $-0.578 < \mathcal I(1) < 0.529$. Thus, despite the opposite median tendencies of the two SN routes, the 95\% credible intervals include zero interaction throughout the redshift range displayed in Figure~\ref{fig:fig3}. The median difference in $E(z)$ between $\xi$IDE and the $\Lambda$CDM reference remains at the few-per-cent level throughout $0<z<5$. This difference reflects correlated changes in $w$, $\xi$, and the present-day dark sector densities, and therefore cannot be attributed solely to the interaction term. The reconstruction provides no statistically significant evidence that nonzero energy transfer is required at any redshift in the range considered. Instead, it reveals a broad degeneracy among interacting solutions whose expansion histories remain close to the $\Lambda$CDM predictions.

\begin{figure}[h]
    \centering
    \includegraphics[width=0.96\textwidth]{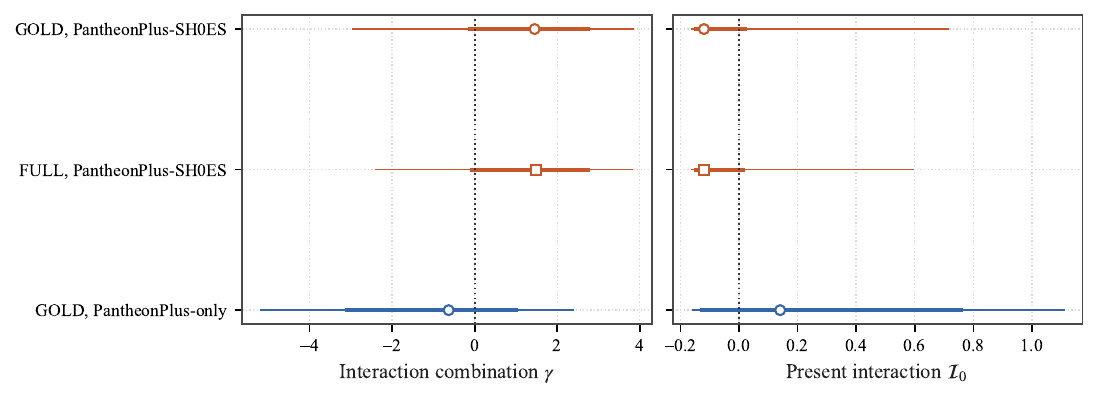}
    \caption{Robustness of the $\xi$IDE inference.  Points show posterior medians of $\gamma$ and $\mathcal I_0$, thick and thin segments show 68\% and 95\% credible intervals. Rows compare the GOLD and FULL GRB samples, and the two SN routes. The within-route changes are small, whereas changing the SN calibration reverses the posterior sign tendency.}
    \label{fig:fig4}
\end{figure}

Figure~\ref{fig:fig4} compares the $\xi$IDE constraints obtained with different GRB samples and SN routes. Replacing GOLD by FULL in the PantheonPlus-SH0ES gives median $\gamma=1.477$, $P(\gamma>0)=0.831$, and $\mathcal I_0=-0.120$. The GOLD and FULL $\xi$IDE posteriors are highly consistent: their 68\% credible intervals overlap strongly for all reported parameters, including $\gamma$, whose 68\% interval contains zero for both samples. The choice of GRB sample does not explain the sign difference. The dominant change is removal of the SN absolute calibration.

The non-interacting models determine whether the fit improvement requires energy exchange. Their PantheonPlus-SH0ES constraints are
\begin{equation}
    w=-0.865^{+0.038}_{-0.038},
    \qquad
    (w_0,w_a)=(-0.772^{+0.067}_{-0.064},-0.800^{+0.425}_{-0.429}),
\end{equation}
while the PantheonPlus-only gives
\begin{equation}
    w=-0.915^{+0.039}_{-0.040},
    \qquad
    (w_0,w_a)=(-0.890^{+0.069}_{-0.058},-0.067^{+0.634}_{-0.478}).
\end{equation}

\begin{figure}[h]
    \centering
    \includegraphics[width=0.94\textwidth]{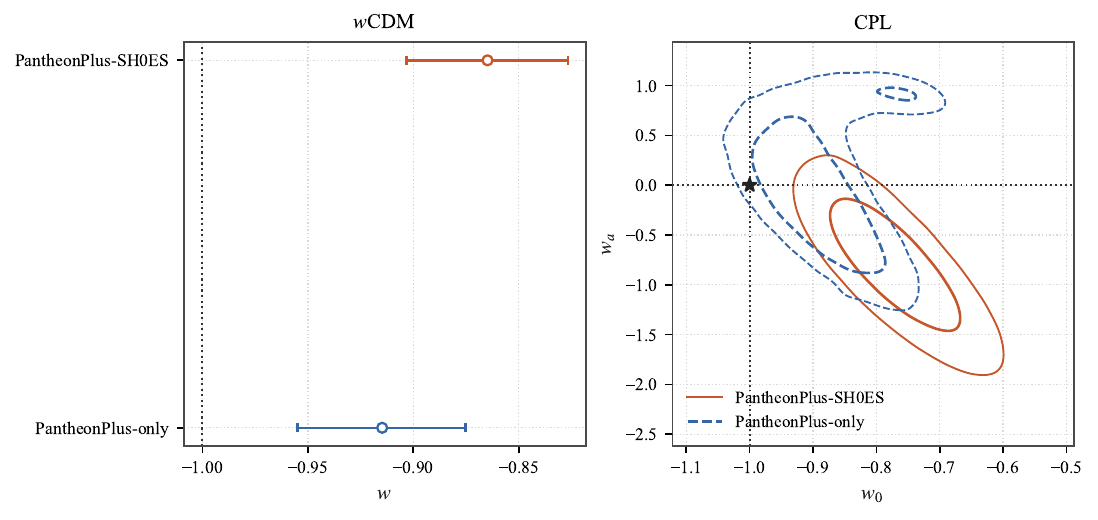}
    \caption{Constraints on the non-interacting models. Left: 68\% intervals for constant $w$, with $w=-1$ marked by the dotted line. Right: 68\% and 95\% CPL contours, the star marks $(w_0,w_a)=(-1,0)$.  The route dependence of these models shows that a changed late-time expansion history is not unique to $\xi$IDE.}
    \label{fig:fig5}
\end{figure}

Figure~\ref{fig:fig5} shows that calibration-dependent flexibility is already present without an interaction. As shown in Table~\ref{tab:table4}, relative to $\Lambda$CDM, the PantheonPlus-SH0ES $\xi$IDE fit has $\Delta{\rm AIC}=-11.520$ and $\Delta{\rm BIC}=-0.453$. However, relative to $w$CDM itself, adding the interaction parameter changes these criteria by only $-1.489$ and $+4.045$, respectively. In the PantheonPlus-only, the corresponding changes are $+1.803$ and $+7.275$. CPL and $\xi$IDE attain nearly identical maximum likelihoods, their $-2\ln\mathcal L_{\max}$ values differ by only 0.166 in the PantheonPlus-SH0ES and 0.050 in the PantheonPlus-only.

\begin{table}[t]
\centering
\caption{Results of model comparison. The $\Delta \mathrm{AIC}$ and $\Delta \mathrm{BIC}$ are relative to $\Lambda$CDM.}
\label{tab:table4}
\begin{tabular}{@{}clrrrr@{}}
\toprule
SN routes
& Models
& $-2\ln\mathcal{L}_{\max}$
& $\Delta{\rm AIC}$
& $\Delta{\rm BIC}$ \\
\midrule

\multirow[c]{4}{*}{PantheonPlus-SH0ES}
& $\Lambda$CDM
& 1782.474
& --
& -- \\

& $w$CDM
& 1770.443
& $-10.031$
& $-4.498$ \\

& CPL
& 1766.788
& $-11.686$
& $-0.620$ \\

& $\xi$IDE
& 1766.954
& $-11.520$
& $-0.453$ \\

\addlinespace
\midrule
\addlinespace

\multirow[c]{4}{*}{PantheonPlus-only}
& $\Lambda$CDM
& 1419.786
& --
& -- \\

& $w$CDM
& 1415.268
& $-2.518$
& $+2.954$ \\
& CPL
& 1415.121
& $-0.665$
& $+10.279$ \\

& $\xi$IDE
& 1415.071
& $-0.715$
& $+10.229$ \\

\bottomrule
\end{tabular}
\end{table}

\section{Summary and Conclusions}\label{sec:sum_con}

We investigate how the treatment of the absolute calibration of SNe Ia affects the constraints on the $\xi$IDE model. The Amati relation and cosmological parameters are fitted simultaneously using \textit{Fermi} GRB measurements together with PantheonPlus SNe Ia, DESI DR2 BAO, and the updated CC compilation, supplemented by a Planck prior to $\omega_b$. The BAO ruler is treated as a free parameter. We used two SN routes, a PantheonPlus-SH0ES that retains the SN absolute calibration and a PantheonPlus-only in which the nuisance SN absolute-magnitude is analytically marginalized.

Neither route excludes the non-interacting limit $\gamma=0$ at 68\% credibility. Although their posterior medians favor opposite directions of energy transfer, the 95\% credible bands for the reconstructed interaction rate $\mathcal I(z)$ include zero throughout the redshift range considered. The inferred sign of interaction is not stable against the treatment of the SN absolute calibration. The results of our PantheonPlus-SH0ES constraints are broadly consistent with Figueruelo et al.\cite{Figueruelo2026}, showing the same tendency towards a low $\Omega_{c0}$ and energy transfer from dark matter to dark energy. The PantheonPlus-only instead favors the opposite direction.

Comparable likelihood gains in $w$CDM and CPL show that the improvement over $\Lambda$CDM is not specific to dark sector interaction. The information criteria do not consistently favor $\xi$IDE, while its reconstructed expansion histories remain close to the route-matched $\Lambda$CDM histories. Constraints on $H_0$ and $r_d$ should be interpreted as conditional on the adopted calibration route and scale prior.

The simultaneous fit to the GRB measurements and the Amati relation avoids constructing a GRB Hubble diagram under an externally selected cosmology. The GOLD and FULL samples yield consistent cosmological constraints, and the redshift evolution diagnostic does not find a significant redshift trend in the Amati slope. Trends in the intercept and intrinsic scatter characterize the observed samples, but they cannot establish intrinsic evolution without an explicit selection model. Overall, although the $\xi$IDE model is consistent with the combined data, we find no compelling evidence for a dark sector interaction that is robust to the choice of SN calibration or specifically favored over noninteracting dark energy extensions.

\section*{Acknowledgments}

This work made use of publicly available data products from the \textit{Fermi} Gamma-ray Burst Monitor, the PantheonPlus SNe Ia compilation, and the DESI DR2 BAO release, together with cosmic-chronometer measurements compiled from the literature. We acknowledge the teams and collaborations responsible for obtaining, validating, and publicly releasing these data. This work was supported by the National SKA Program of China (Grants Nos. 2022SKA0110200 and 2022SKA0110203). This work was also supported by Xiaofeng Yang's ZHISHAN Distinguished Professor startup funding of Henan University.

\bibliographystyle{JHEP}
\bibliography{biblio}

\appendix
\section{Joint constraints on $H_0$ and $r_d$}
\label{app:h0-rd}

\begin{figure}[htbp]
    \centering
    \includegraphics[width=0.95\textwidth]
    {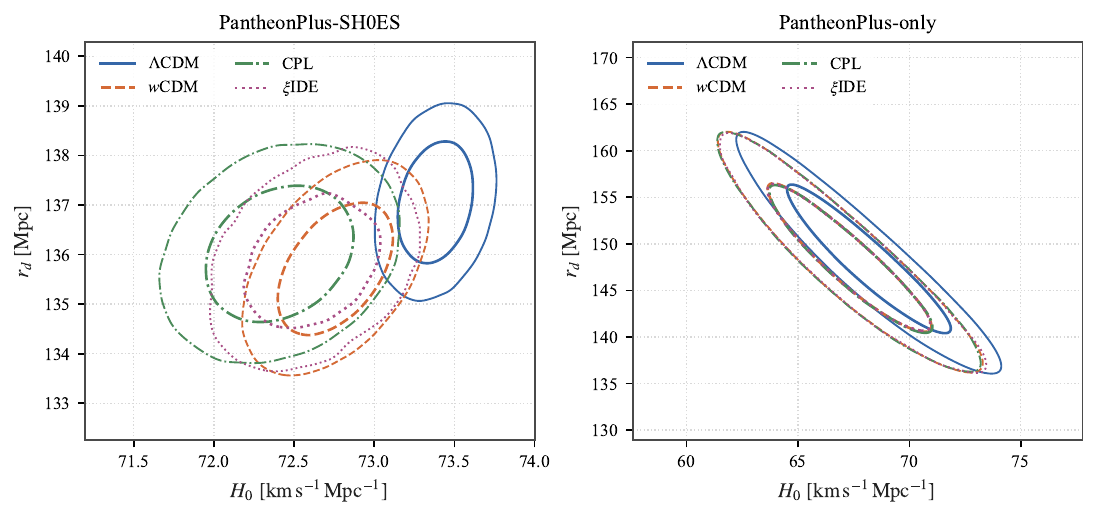}
    \caption{Joint constraints (68\% and 95\% credible contours) in the $H_0$--$r_d$ plane for $\Lambda$CDM, $w$CDM, CPL, and $\xi$IDE. Both panels use the GOLD GRB sample. The left panel uses the PantheonPlus-SH0ES, while the right panel uses the PantheonPlus-only.}
    \label{fig:h0-rd}
\end{figure}

The two SN routes also produce anticorrelated shifts in $H_0$ and $r_d$. For $\xi$IDE, the PantheonPlus-SH0ES gives $H_0=72.61^{+0.27}_{-0.26}\ {\rm km\,s^{-1}\,Mpc^{-1}}$ and $r_d=135.85^{+0.88}_{-0.87}\ {\rm Mpc}$, the PantheonPlus-only gives $H_0=67.27^{+2.36}_{-2.37}\ {\rm km\,s^{-1}\,Mpc^{-1}}$ and $r_d=148.25^{+5.31}_{-4.95}\ {\rm Mpc}$. The values of the more tightly constrained product are $hr_d=98.64^{+0.83}_{-0.82}\ {\rm Mpc}$ and $99.72^{+0.87}_{-0.86}\ {\rm Mpc}$, respectively. Figure~\ref{fig:h0-rd} shows the joint posterior constraints on $H_0$ and $r_d$ and illustrates how the inferred absolute scale depends on the treatment of the SN calibration. Both panels use the GOLD GRB sample.

The PantheonPlus-SH0ES route favors larger $H_0$ and smaller $r_d$, whereas analytic marginalization over the SN absolute-magnitude shifts the posterior toward smaller $H_0$ and larger $r_d$. The contours move mainly along the familiar BAO scale degeneracy. Despite the shifts in the individual parameters, the inferred values of $hr_d$ remain similar between the two routes.

The same displacement is present in $\Lambda$CDM, $w$CDM, CPL, and $\xi$IDE. It is a general consequence of the absolute-scale treatment rather than a feature unique to
the interacting model.

\end{document}